\documentclass[11pt,a4paper]{article}
\usepackage[hyperref]{acl2021}
\usepackage{times}
\usepackage{latexsym}

\usepackage{microtype}
\usepackage{float}
\usepackage{pgfplots}
\pgfplotsset{width=15cm, compat=1.4}
\usetikzlibrary{patterns}
\usepackage{multirow}
\usepackage{amsmath}
\usepackage{array,etoolbox}
\preto\tabular{\setcounter{magicrownumbers}{0}}
\newcounter{magicrownumbers}

\aclfinalcopy 
\def\aclpaperid{24} 

\title{Multi-accent Speech Separation with One Shot Learning}

\author{Kuan Po Huang$^{1\star}$, Yuan-Kuei Wu$^{2\star}$, Hung-yi Lee$^{3}$ \\
  $^{123}$National Taiwan University\\
  $^{1}$Graduate Institute of Computer Science and Information Engineering \\
  $^{23}$Graduate Institute of Communication Engineering \\
  \texttt{\{r09922005, f07942100, hungyilee\}@ntu.edu.tw}
  \thanks{$^\star$The two first authors made equal contributions.}
  }

\date{}

\begin{document}
\maketitle
\begin{abstract}
Speech separation is a problem in the field of speech processing that has been studied in full swing recently. However, there has not been much work studying a multi-accent speech separation scenario. Unseen speakers with new accents and noise aroused the domain mismatch problem which cannot be easily solved by conventional joint training methods. Thus, we applied MAML and FOMAML to tackle this problem and obtained higher average Si-SNRi values than joint training on almost all the unseen accents. This proved that these two methods do have the ability to generate well-trained parameters for adapting to speech mixtures of new speakers and accents. Furthermore, we found out that FOMAML obtains similar performance compared to MAML while saving a lot of time.
\end{abstract}

\section{Introduction}

Speech separation has been a well-known task to solve in the speech processing field. Many model architectures mentioned in Section \ref{sec:related} have been proposed and achieved high performance. This suggests that deep learning based methods are suitable for the speech separation task. 

Despite having promising results, the generalizability of these models is still questionable. The performance of switching to different datasets or environments is not guaranteed. A straightforward solution is to exhaustively collect data under all kinds of environment settings and train a model with these data jointly. Although this may sound reasonable, it is difficult to always consider every situation during training. To make sure that models can be quickly adapted to mixtures spoken by new speakers with not many samples, meta-learning comes to the rescue.  Meta-learning has been widely applied on different speech tasks, especially on speech recognition mentioned in Section \ref{sec:related}. Nonetheless, there is not much work that applied meta-learning on the speech separation task. In our previous work, \citep{wu2020one}, we first proposed to solve the speech separation problem with meta-learning. Their setting is viewing utterance mixtures of two different speakers as a meta task. These speakers have the same accents. However, we hope that a speech separation model can have the ability to adapt to mixtures with accents never seen before. Thus, besides the setting of two different speakers forming a meta task, we also added a setting that meta tasks with speakers of same accents form an accent task set. Section \ref{sec:MAML} and \ref{sec:dataset} describe more about the dataset and task construction procedure.

Our contributions are listed below:
\begin{itemize}
    \item To our best knowledge, we are the first to conduct speech separation experiments on a multi-accent dataset.
    \item We applied meta-learning to help improve the multi-accent speech recognition task.
\end{itemize}

The remaining sections of this paper are organized as follows. In Section \ref{sec:related}, we give a brief overview of existing works related to speech separation and meta-learning. In Section \ref{sec:ss}, we elaborate the problem formulation of speech separation in detail. In Section \ref{sec:MAML}, we list out the two phases of MAML, including the meta training phase and meta testing phase. Additionally, we show how FOMAML is modified from MAML. The experimental setup, dataset, and model we used are presented in Section \ref{sec:exp}. Finally, results and conclusions are given in Section \ref{sec:results} and \ref{sec:conclusion}.

\begin{figure*}[t]
    \centering
    \includegraphics[width=12cm]{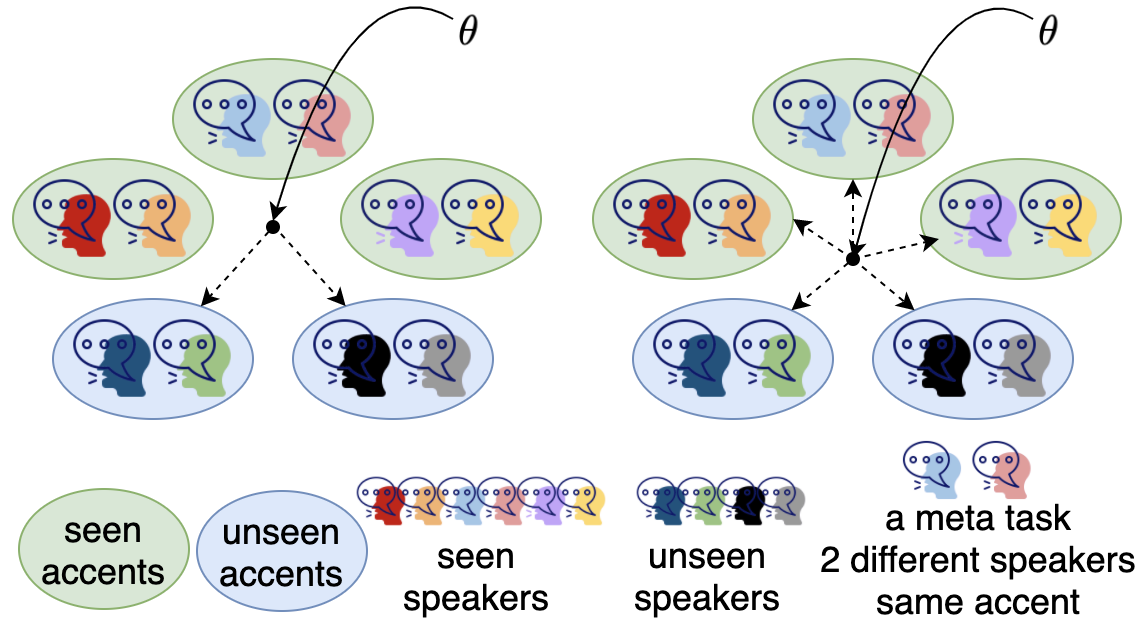}
    \caption{Illustration of joint training and meta-learning for multi-accent speech separation. The oval area is the accent task sets. Each accent task set contains multiple meta tasks. The solid lines are the pretraining process, joint training on the left, and meta-learning on the right. The dashed lines represent the adaptation paths from parameters $\theta$ to the unseen accents of unseen speakers. This figure is modified from \citet{gu2018meta} and our previous work \citet{wu2020one}.}
    \label{fig:joint_meta}
\end{figure*}

\section{Related Work} \label{sec:related}
\paragraph{Speech Separation}
End-to-end separation models have shown great success in separating speech mixtures of the WSJ0-2mix dataset designed by \citep{hershey2016deep} which is generated from the WSJ0 corpus\citep{paul1992design}. \citep{luo2018tasnet} came up with a time-domain audio separation network (TasNet) that takes waveforms as input to alleviate the separation model from dealing with time-frequency representations. They further proposed convolutional TasNet \citep{luo2019conv} which substitutes the LSTM layers in TasNet with convolutional layers. This overcame the problem of long temporal dependencies of LSTM and reduced the model size. Before long, they came up will the Dual-path RNN model, which used intra- and inter-blocks to capture local and global information dependencies within the speech mixtures. \cite{nachmani2020voice} utilized the idea of Dual-path RNN and added a speaker identity loss to improve performance on separating mixtures with an unknown number of speakers. \citep{tzinis2020sudo} proposed to use a separator constructed with U-ConvBlocks which can not only reduce the number of layers while still having high performance but also require less computational resources and time. This helped the model to more likely be used in real-time speech separation. \citep{zeghidour2020wavesplit} integrated speaker identity information into the separating process, and obtained state-of-the-art performance.

\paragraph{Meta-learning} 
Meta-learning has recently become a trend when it comes to solving multi-task problems. This training method has been widely applied in the computer vision field, for instance, \citep{vinyals2016matching, rusu2018meta, sun2019meta}. Meta-learning is also used in the natural language processing field. \citep{gu2018meta} used MAML \citep{finn2017model} for low-resource neural machine translation (NMT). Moreover, in the speech processing domain, some speech-related problems are solved with meta-learning, too. \citep{winata2020meta} applied meta-transfer learning on code-switched speech recognition. \citep{xiao2020adversarial, hsu2020meta} applied meta-learning to solve the multilingual low-resource speech recognition problem. \citep{winata-etal-2019-learning} also used MAML to adapt models to unseen accents on speech recognition. \citep{indurthi2019data} adopted meta-learning algorithms to perform speech translation on speech-transcript paired low-resource data. \citep{chen2021improved} came up with some improvements of meta-learning to help the speaker verification task. 

\section{Speech Separation} \label{sec:ss}
In this work, we perform single channel speech separation. Given a mixture
\begin{equation}
    \mathbf{x} = \sum_{c=1}^C \mathbf{s}_c
\end{equation}
where $C$ is the number of speakers in mixture $\mathbf{x} \in \mathbf{R}^{T}$ and $\mathbf{s}_c \in \mathbf{R}^{T}$ are the ground truth sources. For speech separation, the goal is to estimate $C$ sources $\{\mathbf{\hat{s}}_1, \cdots, \mathbf{\hat{s}}_C\} \in \mathbf{R}^{T}$ such that the estimates sources are as similar as the ground truth sources. The model we used in this work is Conv-TasNet \citep{luo2019conv}. In their work, the similarity of the estimated sources and ground truth sources are measured by scale-invariant signal-to-noise ratio (Si-SNR) shown in Eq.(\ref{eqn:sisnr}):
\begin{equation}
    \mathbf{s}_{\text{proj}} = \frac{\mathbf{s} \cdot \mathbf{\hat{s}}}{\|\mathbf{s}\|^2} \mathbf{s}
\end{equation}
\begin{equation}
    \text{error} = \mathbf{\hat{s}} - \mathbf{s}_{\text{proj}}
\end{equation}
\begin{equation}
    \text{Si-SNR} = 10 \log_{10} \frac{\|\mathbf{s}_{\text{proj}}\|^2}{\|\text{error}\|^2}
    \label{eqn:sisnr}
\end{equation}


The Conv-TasNet model is a mask-based model which consists of an encoder, separator, and decoder. The encoder encodes the mixture $\mathbf{x}$ to a latent space as shown in Eq.(\ref{eqn:encoder}).
\begin{equation}
    \mathbf{x}_{\text{enc}} = \text{enc}(\mathbf{x})
    \label{eqn:encoder}
\end{equation}
$\mathbf{x}_{\text{enc}} \in \mathbf{R}^{H\times T^\prime}$ is the encoder output, where $H$ is the dimension of the latent space and $T^\prime$ is the length of $\mathbf{x}_{\text{enc}}$. The separator then calculates $C$ masks $\mathbf{m}_i \in \mathbf{R}^{H\times T^\prime}$, $i \in \{1, \cdots, C\}$ based on $\mathbf{x}_{\text{enc}}$ shown in Eq.(\ref{eqn:separator}).
\begin{equation}
    \mathbf{m}_i = \text{sep}(\mathbf{x}_{\text{enc}})
    \label{eqn:separator}
\end{equation}
The masks are then multiplied with the encoder output, forming separated features $\mathbf{d}_i$ shown in Eq.(\ref{eqn:mask_output}), 
\begin{equation}
    \mathbf{d}_i = \mathbf{x}_{\text{enc}} \odot \mathbf{m}_i
    \label{eqn:mask_output}
\end{equation}
where $\odot$ is the element-wise multiplication. The separated features $\mathbf{d}_i$ can be viewed as source representations, and are further input to a decoder to estimate separated sources shown in Eq.(\ref{eqn:s_hat}).
\begin{equation}
    \mathbf{\hat{s}}_i = \text{dec}(\mathbf{d}_i)
    \label{eqn:s_hat}
\end{equation}
At this point, before measuring the estimated sources with Si-SNR, there is a label permutation problem. An align between $\{\mathbf{\hat{s}}_1, \cdots, \mathbf{\hat{s}}_C\}$ and $\{\mathbf{s}_1, \cdots, \mathbf{s}_C\}$ needs to be decided. We used the utterance-level permutation invariant training(uPIT) method described in \citep{kolbaek2017multitalker} to solve this problem.

\section{MAML} \label{sec:MAML}
The procedure of MAML \citep{finn2017model} is stated as follows. Given a set of multi-accent tasks $\mathcal{T} = \{ \{\mathcal{T}^i_1\}_{i=1}^{tq_1}, \cdots, \{\mathcal{T}^i_K\}_{i=1}^{tq_K} \}$,
where $K$ is the number of accents. $\mathcal{T}_k = \{\mathcal{T}^i_k\}_{i=1}^{tq_k}$ is the accent task set containing tasks only with the $k^{th}$ accent and $tq_k$ denotes the task quantity of the $k^{th}$ accent task set. The set of tasks $\mathcal{T}$ is split into the source task set $\mathcal{T}_{source}$ and the target task set  $\mathcal{T}_{target}$. The model denoted as $f$, will be trained on the source task set $\mathcal{T}_{source}$ in the hope of having the ability to quickly adapt to the target task set $\mathcal{T}_{target}$. 
\subsection{Meta Training Phase}
During the meta training phase, the MAML algorithm aims to find initialized parameters $\theta$ that can further be quickly adapted to new tasks. Moreover, these initialized parameters should be sensitive to the difference between two different tasks, such that adaptation of the initialized parameters can significantly improve the performance on new tasks sampled from the source task set $\mathcal{T}_{source}$. This is achieved by the inner loop and outer loop optimization. A batch of tasks $\tau_{source} = \{ \tau_1, \cdots, \tau_b \}$ is sampled from $\mathcal{T}$ proportional to the task quantity of every accent task set, e.g., for an accent task set $\mathcal{T}_k$, the larger $tq_k$ is, the more likely a task is to be sampled from it. Each task in $\tau_{source}$ is further split into a support set $\tau^{sup}$ and a query set $\tau^{qry}$. The support set is used to adapt the model parameters by performing a one-step gradient decent, which is known as the inner loop shown in Eq.(\ref{eqn:inner_loop}).
\begin{equation}
    \theta_j^\prime \leftarrow \theta - \alpha \nabla_\theta\mathcal{L}_{\tau_j^{sup}}(f_\theta)
    \label{eqn:inner_loop}
\end{equation}
where $\alpha$ is the learning rate. The goal of the inner loop is to minimize the loss of $\tau_j^{sup}$ with respect to $f_{\theta}$. More concisely,
\begin{equation}
    \theta_j^\prime = \arg\min_{\theta} \mathcal{L}_{\tau^{sup}_j}(f_{\theta})
\end{equation}
At this point, the sum of the query loss of each query set in $\tau_{source}$ is calculated by
\begin{equation}
    \mathcal{L}_{qry} = \sum_{j=1}^{b} \mathcal{L}_{\tau^{qry}_j}(f_{\theta_j^\prime})
\end{equation}
The goal of the meta training phase is to minimize the total loss of the query sets. This is also performed by a one-step gradient decent, known as the outer loop shown in Eq.(\ref{eqn:outer_loop}).
\begin{equation}
    \theta \leftarrow \theta - \beta\nabla_\theta\mathcal{L}_{qry}
    \label{eqn:outer_loop}
\end{equation}
\subsection{Meta Testing Phase}
During the meta testing phase, we perform a procedure (see Eq.(\ref{eqn:adapt})) similar to the inner loop in the meta training phase. This procedure adapts the parameters $\theta$ obtained in the meta training phase to the target tasks $\tau_{target} = \{ \tau^\prime_1, \cdots, \tau^\prime_b \}$. 
\begin{equation}
    \theta_j \leftarrow \theta - \beta \nabla_\theta\mathcal{L}_{\tau_j^{\prime sup}}(f_\theta)
    \label{eqn:adapt}
\end{equation}

\subsection{First-order MAML (FOMAML)}
Eq.(\ref{eqn:gradient}) is the calculation of the  gradient in the outer loop, where $\mathcal{L}_{\tau^{qry}_j}$ is denoted as $\mathcal{L}^j$ for simplicity.
\begin{equation}
    \nabla_\theta\mathcal{L}_{qry} = \nabla_\theta\sum_{j=1}^{b} \mathcal{L}^j(f_{\theta_j^\prime})
    = \sum_{j=1}^{b}\nabla_\theta \mathcal{L}^j(f_{\theta_j^\prime})
    \label{eqn:gradient}
\end{equation}
When performing the outer loop during the meta training phase, high computational cost is needed to calculate the second-order derivatives with backpropagation. Eq.(\ref{eqn:approx}) is the first-order approximation of the second-order derivative,
\begin{equation}
    \frac{\partial \mathcal{L}^j(f_{\theta_j^\prime})}{\partial \theta^d} 
    = \sum_{i=1}^D \frac{\partial \mathcal{L}^j(f_{\theta_j^\prime})}{\partial \theta_j^{\prime i}}\frac{\partial \theta_j^{\prime i}}{\partial \theta^d} 
    \approx \frac{\partial \mathcal{L}^j(f_{\theta_j^\prime})}{\partial \theta_j^{\prime d}}
    \label{eqn:approx}
\end{equation}
where $\theta$ is a $D$ dimensional parameter, $\theta^d$ is the $d$-th dimension of $\theta$ and $\theta_j^{\prime i}$ is the $i$-th dimension of $\theta_j^{\prime}$. The difference between FOMAML and MAML is that this approximation is used instead of the second-order derivatives. Thus, compared to MAML, FOMAML can save a lot of computational time, resulting in a faster gradient calculation.

\section{Experiments} \label{sec:exp}
\subsection{Dataset}\label{sec:dataset}
\begin{figure}[t]
    \centering
    \includegraphics[width=6cm]{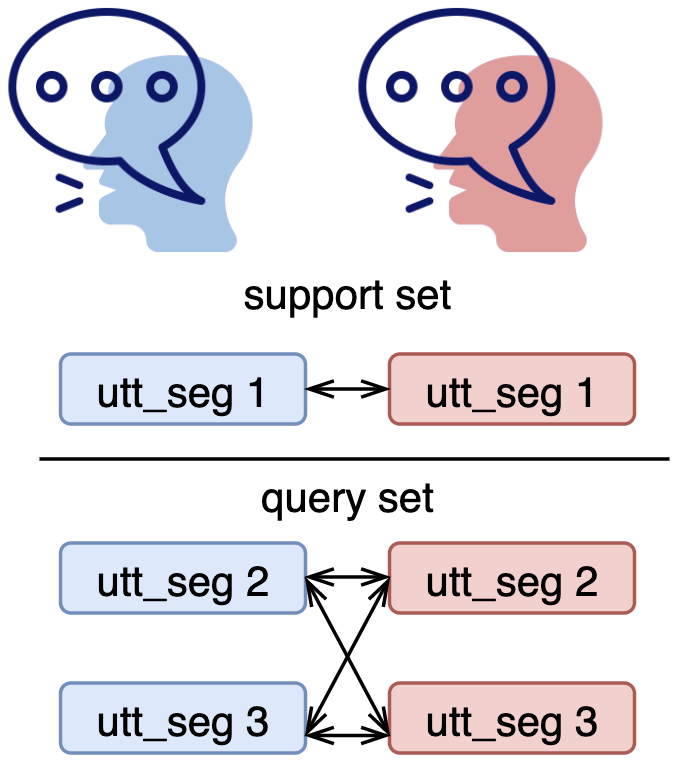}
    \caption{Illustration of a meta task. For two different speakers with the same accent, we sample 3 utterance segments to form a meta task. Thus, there will be 9 mixtures. However, during training, we only sample one mixture to form the support set since our setting is one shot learning. The other 4 mixtures that do not contain the utterance segments in the support set are selected to form the query set.}
    \label{fig:meta_task}
\end{figure}
The multi-accent speech utterances are collected from the speech accent archive \citep{Wei2014speech}. This archive currently has more than 200 kinds of accents and 2939 samples. Each native or non-native speaker speaks the same English paragraph. We selected 123 accents that contain more than one speaker since we need utterances of two different speakers to generate mixtures. We split these accents into three sets, 85 accents for generating the training tasks and 19 accents each for generating the developing and testing tasks. The utterance of each speaker is split into segments with a duration of 4 seconds. For each accent, we construct meta tasks by following the task construction method described in \citep{wu2020one}. We select at most 12 speakers for each accent and generate speech mixtures for each pair of speakers with the same accents. Thus, there will be at most ${12 \choose 2} = 66$ meta tasks and at least ${2 \choose 2} = 1$ meta task for each accent. In each meta task, 3 utterance segments are selected from each speaker and mixed with an SNR level randomly selected between 0 to 5 dB and resampled at an 8kHz sample rate. This results in $3\times 3=9$ speech mixtures in one meta task. Fig.(\ref{fig:meta_task}) is an illustration describing the support set and query set of a meta task. Finally, for the training, developing, and testing set, 22.4, 3.8, and 3.9 hours of speech mixtures are generated.

\subsection{Model}
The model we used is Conv-TasNet \citep{luo2019conv}. It consists of an encoder, separator, and a decoder. The encoder is a 1-dim convolution, which transforms the input mixture into a representation. The separator then calculates two masks based on the encoder output. More specifically, it consists of $R$ stacks of temporal convolutional networks (TCN). Each TCN layer consists of $M$ 1-dim exponentially increasing dilated convolutional blocks. These $M$ blocks each have a residual connection and a skip connection. The residual connection is the input of the next block and the skip connection of all blocks are summed together, passing a parametric relu, linear projection, and a sigmoid function to produce two masks. The two masks are multiplied with the representation output from the encoder respectively and further input into the decoder to generate two separate waveforms of the two speakers. The decoder is also a 1-dim convolution. The configuration that we used is the one that obtained the best performance reported in \citep{luo2019conv}.

\begin{figure}[ht]
    \centering
    \begin{tikzpicture}
\begin{axis}[
    width=.50\textwidth,
    height=6.0cm,
    xlabel={$\beta$},
    ylabel={Si-SNRi},
    xmin=0.5, xmax=10.5,
    ymin=-3, ymax=11,
    xtick={1,...,9},
    xticklabels={1e-5, 5e-5, 1e-4, 5e-4, 1e-3, 5e-3, 1e-2, 5e-2, 1e-1},
    x tick label style={rotate=40,anchor=east},
    ytick={-3,...,10},
    yticklabels={, -2, , 0, , 2, , 4, , 6, , 8, , 10},
    legend style={nodes={scale=0.8, transform shape}}, 
    legend pos=north east,
    ymajorgrids=true,
    grid style=dashed,
    label style={font=\small},
    tick label style={font=\small} 
]

\addplot[
    color=red,
    mark=square,
    mark options={scale=0.5}
    ]
    coordinates {
        (1, 8.44)
        (2, 8.46)
        (3, 8.48)
        (4, 8.52)
        (5, 8.47)
        (6, 7.83)
        (7, 6.92)
        (8, 1.32)
        (9, -2.25)
    };
\addplot[
    color=blue,
    mark=square,
    mark options={scale=0.5}
    ]
    coordinates {
        (1, 6.71)
        (2, 6.79)
        (3, 6.72)
        (4, 6.86)
        (5, 6.89)
        (6, 6.66)
        (7, 5.93)
        (8, 1.09)
        (9, -2.59)
    };
    \legend{w/o noise, w/ noise}
    
\end{axis}
\end{tikzpicture}
    \caption{For fine-tuning after joint training, we evaluated the performance by adjusting the learning rate $\beta$ in the range of $10^{-5}$ to $10^{-1}$.}
    \label{fig:fastlr}
\end{figure}
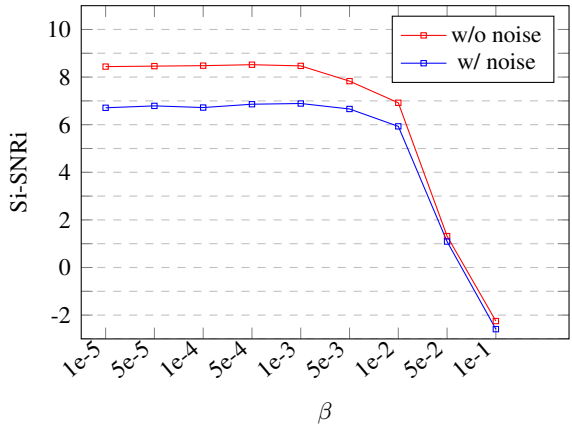
\begin{table*}[ht]
    \centering
    \begin{tabular}{rlc|c|c}
    \hline
    & \textbf{method} & \textbf{fine-tune} & \textbf{test w/o noise} & \textbf{test w/ noise}\\ \hline
    (a)& & before & 8.40 $\pm$ 2.25 & 6.67 $\pm$ 2.10 \\
    (b)& \multirow{-2}{*}{Joint Training} & after & 8.52 $\pm$ 2.20 & 6.89 $\pm$ 1.84 \\ \hline
    (c)& & before & 8.45 $\pm$ 3.19 & 6.66 $\pm$ 2.59\\ 
    (d)& \multirow{-2}{*}{FOMAML} & after & \textbf{10.13} $\pm$ 2.12 & 8.19 $\pm$ 1.62\\ \hline
    (e)& & before & -6.19 $\pm$ 1.38 & -6.85 $\pm$ 1.31 \\
    (f)& \multirow{-2}{*}{MAML} & after & 10.11 $\pm$ 1.86 & \textbf{8.26} $\pm$ 1.52 \\ \hline
\end{tabular}
    \caption{Evaluation results of joint training and MAML methods on the testing accent task sets with and without noise. The two numbers in a cell denote the average Si-SNRi of all the testing tasks and the standard deviation of all the testing accent task sets.}
    \label{tab:base_maml}
\end{table*}
\begin{figure*}[h]
    \centering

%
%
%


\begin{tikzpicture}
\begin{axis}[
    width=1.0\textwidth,
    height=6.2cm,
	x tick label style={rotate=40,anchor=east},
	symbolic x coords={ hausa,lithuanian,bari,quechua,yiddish,kurdish,synthesized,tamil,russian,thai,italian,ewe,mende,malay,basque,albanian,ga,estonian, rotuman, a},
	ylabel=Si-SNRi,
	ymin=0,
	ymax=13,
	minor y tick num = 4,
	enlargelimits=0.03,
	legend style={at={(0.5,1.20)},
		anchor=north, legend columns=-1, style={column sep=0.2cm}},
	ybar interval=0.8,
	grid=both,
	grid style=dashed
]
\addplot [fill=cyan] 
table{
    x y
    hausa 7.797184308
    lithuanian 9.136580769
    bari 8.395817757
    quechua 5.02066803
    yiddish 8.00891211
    kurdish 10.322899
    synthesized 9.448816299
    tamil 11.9287285
    russian 10.78634089
    thai 10.36514382
    italian 10.34953827
    ewe 12.40219498
    mende 9.443446795
    malay 10.45881119
    basque 12.67417431
    albanian 10.20583577
    ga 11.19834736
    estonian 10.53571637
    rotuman 4.925264359
    a 0
};
\addplot [fill=magenta] 
table{
    x y
    hausa	6.490743973
    lithuanian	5.875266059
    bari	5.098831177
    quechua	3.496987581
    yiddish	6.180011159
    kurdish	9.26032373
    synthesized	9.217527628
    tamil	10.25052309
    russian	9.183664954
    thai	9.100280969
    italian	8.237139283
    ewe	9.664775848
    mende	9.2962931
    malay	9.330113029
    basque	11.50267029
    albanian	8.431665901
    ga	8.940525913
    estonian	9.342681646
    rotuman	3.953201056
    a 0
};
\addplot [draw=blue, pattern color = blue, pattern = north west lines] 
table{
    x y
    hausa 5.847688998
    lithuanian 7.762980398
    bari 7.402003288
    quechua 3.395904064
    yiddish 7.511906099
    kurdish 8.725517008
    synthesized 8.44408083
    tamil 9.54634254
    russian 8.603444244
    thai 8.493770471
    italian 8.319708953
    ewe 9.803808212
    mende 8.33370177
    malay 8.489306164
    basque 9.18904686
    albanian 7.833111518
    ga 8.99476579
    estonian 8.435000499
    rotuman 4.684219837
    a 0
};
\addplot [draw=red, pattern color = red, pattern = north west lines] 
table{
    x y
    hausa	5.064400539
    lithuanian	5.143381796
    bari	3.14477253
    quechua	3.829597473
    yiddish	5.576066929
    kurdish	7.490164114
    synthesized	7.241561731
    tamil	8.042828202
    russian	7.501661555
    thai	7.313337986
    italian	6.673049019
    ewe	6.865758419
    mende	5.675171018
    malay	7.413916826
    basque	9.195604324
    albanian	6.697544
    ga	7.921788836
    estonian	7.749934435
    rotuman	2.029411316
    a 0
};

\legend{fomaml\_clean\ ,joint\_clean,\ fomaml\_noise, \ joint\_noise}
\end{axis}
\end{tikzpicture}
    \caption{Evaluation results of each testing accent task set for model (b) and (d) in table \ref{tab:base_maml}.}
    \label{fig:hist}
\end{figure*}
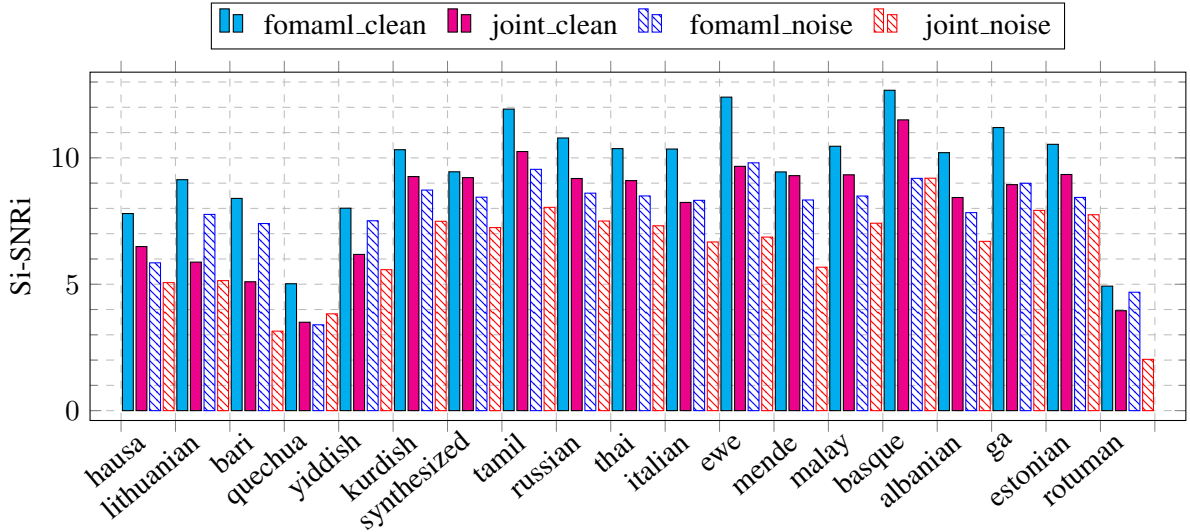

\subsection{Joint Training and Transfer Learning}
There are many other works such as \citep{chen2020darts, tong2017investigation}, that try to solve the domain mismatch problem, where the source domain and target domain datasets do not have a similar distribution. Joint training refers to pretraining a model with different source domain data together. Transfer learning refers to adapting the pretrained model to some partial target domain data and testing the fine-tuned model on the target domain data. The most common adaptation method is fine-tuning. Moreover, the domain mismatch scenario has a low-resource problem if the target domain has only fewer data compared to the scale of the source domain data. There are also several works that tried to solve this problem, such as \citep{chen2015multitask, zoph2016transfer, hsu2020meta}. Our jointly trained model is also based on this low-resource scenario.

\subsection{MAML and FOMAML}
To deal with the domain mismatch and low-resource problem, we applied MAML as our training method in the hope of performing better than joint training. We set the number of the support set in each task as 1, meaning that the model needs to have the ability to adapt to a new task by only seeing one speech mixture of two new different speakers with a new accent never seen before. We also trained our model with FOMAML in order to know whether calculating gradients with first-order approximation still obtains relatively good performance compared to training with MAML.

\subsection{Experiment Settings}
For both the joint training and MAML methods, we trained the model from randomly initialized parameters for 100 epochs with the Adam optimizer of $0.001$ learning rate and $0.00001$ weight decay. For the MAML methods, during the meta training phase, we set $\alpha=0.01$. For joint training, we also fine-tuned the model parameters with the method in Eq.(\ref{eqn:adapt}). We tested the fine-tuning learning rate $\beta$ on the testing set, reported it in section \ref{sec:results}, and used the learning rates that obtained the best performance for joint training as our baseline. However, for the models trained with MAML methods, the fine-tuning learning rate $\beta$ is fixed at $0.01$ since other values lead to significant performance degradation.

\section{Results} \label{sec:results}

\subsection{Joint Training}

For joint training, we tested the fine-tuning learning rate $\beta$ on the testing set as shown in Fig.(\ref{fig:fastlr}), and found out that $\beta=5e{-}4$ obtained the best performance on the clean testing set, while $\beta=1e{-}3$ obtained the best performance on the testing set with noise. We use these two experiment settings as our baseline.

\subsection{MAML and FOMAML}

Comparing models (d), (f) with model (b), we can see that MAML and FOMAML perform better than the joint training baseline. This suggests that the initial model parameters obtained by MAML and FOMAML have the better potential to be adapted to new unseen tasks. Besides, the standard deviation of the testing accent task sets of models (d) and (f) are both less than model (b). This implies that the performance of the models trained with MAML and FOMAML have small dispersion with respect to the mean Si-SNRi value of all the accents compared to the model jointly trained. From Fig.(\ref{fig:hist}), we can see that model (d) performs better on all accents when there is no noise involved and performs better on most of the accents when there is noise in the mixtures.

By comparing models (d) and (f), we found out that these two training methods have similar performance. Model (d) has a slightly higher performance than model (f) under the circumstances that the mixtures are clean in the testing tasks, while model (d) has a slightly lower performance than model (f) under the circumstances that there is noise in the testing tasks. However, MAML requires more than 10 times the training time compared to FOMAML, indicating that the first-order approximation takes advantage over calculating the second-order derivatives by saving a lot of time while still obtaining similar performance. Moreover, FOMAML without fine-tuning (model (c)) has similar performance compared to the baseline model, and yet somehow, initialized parameters obtained by MAML (model (e)) do not have the ability to perform speech separation.

\section{Conclusion} \label{sec:conclusion}
Our results show that MAML and FOMAML training methods are effective on multi-accent speech separation. More specifically, it is confirmed that these two methods are better than joint training when adapting to new speakers with new accents and even noisy environments. Besides, FOMAML is shown to be sufficient for dealing with the multi-accent speech separation task and can reduce a large amount of training time. Despite the fact that FOMAML outperforms joint training on the testing set, we can still see that the performance of each accent task set varies a lot from Fig.(\ref{fig:hist}). This is probably due to the task-difficulty imbalance issue described in \citep{xiao2020adversarial}, perhaps some speakers with special accents may be hard to separate. Thus, in the future, we will try to solve this problem with meta sampling methods mentioned in \citep{xiao2020adversarial}.

\section*{Acknowledgments}
We thank to the National Center for High-performance Computing (NCHC) for providing computational resources and storage.

\bibliographystyle{acl_natbib}
\bibliography{anthology,acl2021}

\end{document}